\documentclass{article}
\usepackage{dcase2023_techrep,amsmath,graphicx,url,times,booktabs, tabularx}

\title{LATENT DIFFUSION MODEL BASED FOLEY SOUND GENERATION SYSTEM FOR DCASE CHALLENGE 2023 TASK 7}

\name{Yi Yuan$^{1}$,
      Haohe Liu$^{1}$,
      Xubo Liu$^{1}$, 
      Xiyuan Kang$^{1}$,
      Mark D. Plumbley$^{1}$, 
      Wenwu Wang$^{1}$
      }
\address{$^1$ University of Surrey, Guildford, United Kingdom\\          
 }

\begin{document}

\ninept
\maketitle

\begin{sloppy}

\begin{abstract}
Foley sound presents the background sound for multimedia content and the generation of Foley sound involves computationally modelling sound effects with specialized techniques. In this work, we proposed a system for DCASE 2023 challenge task 7: Foley Sound Synthesis. The proposed system is based on AudioLDM, which is a diffusion-based text-to-audio generation model. To alleviate the data-hungry problem, the system first trained with large-scale datasets and then downstreamed into this DCASE task via transfer learning. Through experiments, we found out that the feature extracted by the encoder can significantly affect the performance of the generation model. Hence, we improve the results by leveraging the input label with related text embedding features obtained by a significant language model, i.e., contrastive language-audio pretraining~(CLAP). In addition, we utilize a filtering strategy to further refine the output, i.e. by selecting the best results from the candidate clips generated in terms of the similarity score between the sound and target labels. The overall system achieves a Fréchet audio distance~(FAD) score of $4.765$ on average among all seven different classes, substantially outperforming the baseline system which performs a FAD score of $9.7$. 
\end{abstract}


\section{Introduction}
\label{sec:intro}

The remarkable breakthroughs of deep learning models have contributed to success in sound generation~\cite{audioldm,diffsound,makeaudio,audiogen}. Foley sounds, on the other hand, play an important role in enhancing the perceived acoustic properties of movies, music, videos and other multimedia content. Hence, the automatic Foley synthesis system holds immense potential in simplifying traditional sound generation processes, such as manual recording and mixing by human artists. 

Currently, most of the sound generation models adopt an encoder-decoder architecture, which has shown remarkable generation performance. The official baseline system of task 7~\cite{Liu-tts} utilizes a conventional neural network~(CNN) encoder, a variational autoencoder~(VAE) decoder and a generative adversarial network~(GAN) vocoder. The encoder encodes the input feature~(e.g., label) into latent variables and the decoder can decode this intermediate information into mel-spectrogram for the vocoder to generate the final waveform. 

This report describes the methods we submitted to Task 7 of DCASE 2023 challenge~\cite{task7}. The task involves synthesizing sounds across seven different classes, including animal sounds ~(e.g., dog barking), machine sounds ~(e.g., moving motor) and natural sounds (e.g., rain). Similar to image generation, sound synthesis systems are usually implemented by generating a mel-spectrogram or waveform~\cite{liu22}, which poses a challenging task when the waveform appears similar structure in the frequency domain(e.g., rain and motor sounds). Moreover, the scarcity of data within each class makes training a system from scratch even more difficult. To address the issue of data scarcity, we follow the idea of pre-training\cite{yuan2023}, by initially train the models on large-scale datasets such as AudioSet~\cite{audioset} and AudioCaps~\cite{audioset}, then transfer them into the task development set. Our models are primarily based on AudioLDM~\cite{audioldm}, an audio generation model that comprises a diffusion encoder, a VAE decoder and a GAN vocoder. For inputs, the category labels are given into a contrastive language-audio pre-training~(CLAP)~\cite{clap} for input embeddings. We conduct studies on different combinations of the label and texts and leverage the label with text embeddings that can present more useful information. For outputs, we apply a cosine-similarity score between the generated sounds and target labels as a filtering strategy, selecting the most relevant sounds to enhance the overall quality of the final outputs. Through experiments with different sizes of the LDM model and pre-trained CLAP, we observed that generating more complex sounds (e.g., motor and rain) with a larger system leads to lower Fréchet audio distance~(FAD) scores in the validation set. To achieve better overall results, our proposed system ensembles two networks for generating different sound classes. Compare with the baseline system with an average FAD of $9.7$, our system significantly improves by a large margin, achieving a FAD of $4.765$. 

The subsequent sections of this technical report are structured as follows: Section 2 provides an overview of the proposed system. The methodology employed by the network is detailed in Section 3. Section 4 presents the experimental setup utilized. Results are presented in Section 5. Finally, Section 6 summarizes this work and draws conclusions.

\section{SYSTEM OVERVIEW}
\label{sec:format}

Similar to the baseline system, he proposed system adopts a commonly employed architecture for sound generation, comprising an encoder, a generator, a decoder, and a vocoder. Our system follows the same structure as AudioLDM, which incorporates a pre-trained audio-text embedding model~\cite{clap} as the encoder and utilizes a latent diffusion-based model as the generator.

\begin{figure*}[htbp]
    \centering
    \includegraphics[width=\linewidth]{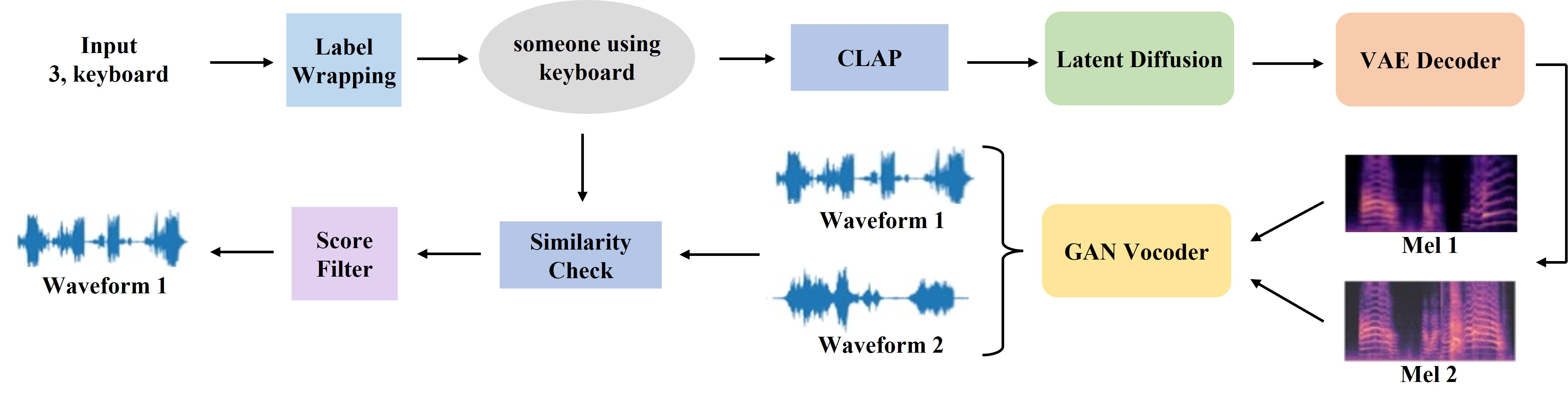}
    \caption{The overview of the system}
    \label{fig:overview}
\end{figure*}

Instead of directly using labels as the input, we employ a text description for each label as the input for the system, such as text: “someone using keyboard”, for label: “3, keyboard”. In our model, the decoder and vocoder undergo separate training processes. Once trained, these two models are integrated into the overall system with fixed parameters. By utilizing the text feature extraction from CLAP as a conditioning input, the LDM generates intermediate sound representations as vectors in the latent space. Afterwards, the VAE decoder decodes these representations into mel-spectrogram results, which are then further reconstructed into waveforms by the GAN vocoder. This system is then further improved with two techniques. First, transfer learning is introduced to boost performance by pre-training the model on larger datasets. Second, a similarity score has been applied after each generation to select only the best match results. Detailed explanations of these methods are provided in the following section. 
The overall sampling procedure of the system is shown in Figure \ref{fig:overview}

\section{METHODOLOGY}

\subsection{Embedding encoder}
For sound generation, we utilized the Contrastive Language-Audio Pretraining (CLAP) model to generate input embeddings. CLAP consists of a text encoder $\textit{f}{text}$, which converts text descriptions \textit{y} into text embeddings $\boldsymbol{E}^{y}$, and an audio encoder, denoted as $\textit{f}{audio}$, which computes audio embeddings, $\boldsymbol{E}^{x}$, from audio samples, represented as \textit{x}. Both encoders are trained using cross-entropy loss on extensive datasets, resulting in a latent space with the same dimensionality for both audio and text embeddings. Leveraging the cross-modal information obtained from the two encoders, we pre-trained our system on larger datasets using audio embeddings, and subsequently fine-tuned it on the smaller-scale task development set using text embeddings.

\subsection{Diffusion generator}
Our system applied a latent diffusion model~(LDM) that takes the feature embedding as the condition and generates the intermediate latent tokens for the decoder. LDM consists of two processes. The forward process involves incrementally adding noise $\boldsymbol{\epsilon}$ to the latent vector $\boldsymbol{z}{0}$, resulting in a sequence of latent vectors $\boldsymbol{z}{n}$ over $\textit{N}$ steps. Then, the reverse process entails the model predicting the transition probabilities $\boldsymbol{\epsilon}{\theta}$ for each step \textit{n}. This allows the denoising process of the noisy latent vector $\boldsymbol{z}{n}$ to transform back into the original data.
During training, the model is trained with a re-weighted objective~\cite{DDPM} as:

\begin{equation}
    L_{n}(\theta)={E}_{\boldsymbol{z}_{0},\boldsymbol{\epsilon},n}|| \boldsymbol{\epsilon} - \boldsymbol{\epsilon}_{\theta}(\boldsymbol{z}_{n},n,\boldsymbol{E}^{x})||^2_{2}
\end{equation}

During the sampling process, the model generates a result \textit{x} by utilizing a sample of Gaussian noise $\textit{x}_{0}$ along with the reverse transition probability learned during training and the text condition $\boldsymbol{E}^{y}$ from CLAP is incorporated. In training, we set the number of denoising steps \textit{N} as $1000$, but during sampling, we only consider $200$ steps.

\subsection{VAE decoder \& HiFi-GAN vocoder}
We conducted training of a Variational Autoencoder~(VAE) to decode the latent feature tokens into mel-spectrograms. During the training process, the VAE learns to compress the mel-spectrograms, denoted as $\boldsymbol{X}$, into a latent space vector $\boldsymbol{z}$ with a compression level of $8$. It then reconstructs the mel-spectrograms back into $\hat{\boldsymbol{X}}$. As for the vocoder, we utilized a HiFi-GAN to generate the sound waveform, represented as $\hat{\textit{x}}$, from the reconstructed mel-spectrograms $\hat{\boldsymbol{X}}$.

\subsection{Transfer learning}
External data is allowed in this task, which allows transfer learning to be used. In our system, all three models adapt the transfer learning. The LDM model undergoes initial training on extensive datasets using audio embeddings as input. Subsequently, this model is fine-tuned on our specific task dataset by utilizing text embeddings.

\subsection{Similarity selection}
To enhance the sound quality further, we incorporate a scoring mechanism into the system to identify the most suitable results. Leveraging the fact that CLAP provides embeddings in a shared latent space for both audio and text, we employ cosine-similarity to measure the relevance between the generated audio and the target text. Through experiments involving different score thresholds, we establish specific thresholds for each sound class. These thresholds enable the system to select only the results that exceed the designated thresholds. Additionally, we observe that the sound of a motor encompasses a combination of noise and engine sounds, resulting in significant diversity and a noticeable distinction between the text embedding and the target sound embedding. To further enhance the filtering function for the motor class, we employ Fréchet audio distance~(FAD). This enables the model to select several audio embeddings from the training set that best matches the motor class and calculate the similarity score between the output audio embedding and the target embedding. A comprehensive comparative analysis of the results, including the normal audio-text approach, is presented in Table~\ref{table:motor} within the results section.

\begin{table*}[htbp]
\centering
\begin{tabular}[\linewidth]{cccccccc}
\hline
\multicolumn{1}{c}{System}        & Dog Bark      & Footstep     & Gun Shot       & Keyboard     & Moving Motor Vehicle & Rain          & Sneeze Cough  \\ \hline
\multicolumn{1}{c}{Basline~\cite{Liu-tts}}       & $13.41$        & $8.11$         & $7.95$          & $5.23$        & $16.11$               & $13.34$        & $3.77$          \\
\multicolumn{1}{c}{LDM\_S\_label} & $4.17$          & $6.86$          & $7.25$           & $3.15$         & $15.68$                & $12.95$         & $2.85$          \\
\multicolumn{1}{c}{LDM\_S\_text}  & $3.84$          & $5.66$          & $6.66$           & $3.48$         & $14.35$                & 12.62         & 2.12          \\ 
LDM\_S\_filter                     & $\textbf{3.53}$ & $\textbf{5.04}$ & $\textbf{5.655}$ & $\textbf{2.8}$ & $15.29$                & $9.76$          & $\textbf{1.92}$ \\
LDM\_L\_label                      & $9.99$          & $7.26$          & $6.83$           & $3.45$         & $13.71$                & $6.81$          & $3.45$          \\
LDM\_L\_text                       & $8.47$          & $8.87$          & $6.75$           & $2.84$         & $13.14$                & $6.16$          & $3.02$          \\
LDM\_L\_filter                     & $6.73$          & $5.15$          & $6.69$           & $2.98$         & $\textbf{12.12}$       & $\textbf{5.53}$ & $2.61$  \\
\hline
\end{tabular}
\caption{The results of the two models with different settings, the label indicates the model takes the label as input while text means that the model takes the text information as input. Filter models are text-embedding models with a similarity score filtering strategy and the filters for motor sound are used with the text embedding of “ A moving motor ”.}
\label{table:results}
\end{table*}

\section{EXPERIMENTS}
\subsection{Dataset}
\textbf{Challenge official dataset} provides a training set with seven different sound classes, each class has around $600$ to $800$ 4-second sounds respectively. All the data is provided as a sound-label pair. 

\noindent \textbf{AudioSet} is an extensive audio dataset that encompasses a diverse array of sounds. More specifically, AudioSet offers approximately $2.1$ million 10-second audio clips accompanied by corresponding labels. During the pre-training phase, our system exclusively utilizes AudioSet data.

\noindent \textbf{Freesound} is another audio-label dataset, albeit with variable lengths for the audio clips. In order to ensure consistency in the output length, all the sounds within Freesound are trimmed to match the duration of 10-second clips.

Combining AudioSet and Freesound, we collected around $2.2$M sounds for pre-training the LDM, VAE and GAN models, while all these models are then fine-tuned into this task with the official dataset. 

\subsection{Evaluation metrics}
We follow the official guidance and apply the Fréchet audio distance~(FAD) score as our main evaluation metric. Specifically, FAD computes the Fréchet distance between the embedding features of two groups of sounds, which are extracted using VGGish~\cite{vggish}. A lower FAD indicates higher audio quality, as it signifies a closer similarity between the generated audio and the target audio.

\subsection{Experimental setup}

As an ensemble model, the decoder and vocoder undergo independent training processes. Subsequently, these two models are integrated into the overall system with fixed parameters to train the Latent Diffusion Model (LDM). Initially, all the models are pretrained from scratch using AudioSet and Freesound datasets, then further fine-tuned using the development set.

In the case of the model LDM\_S, we utilize the mel-spectrogram with a frequency of 22kHz as the input and set the VAE compression level to 4. On the other hand, for the larger LDM model with a larger CLAP model~(LDM\_L), we train it on 16Hz sounds and subsequently upsample them to 22kHz before generating the output to alleviate the complexity of high-dimension computation. The results obtained from both models are presented in the subsequent section.

\section{RESULTS}
Table~\ref{table:results} presents the performance of our system on the validation set. As indicated by the FAD scores, our models consistently outperform the baseline~\cite{Liu-tts} by a significant margin. Notably, the different sizes of the LDM models exhibit distinct strengths: the smaller model excels in generating distinct sounds like dog barks, footsteps, and gunshots, while the larger model demonstrates superior performance in handling more complex sounds such as motor sounds and rain sounds. Additionally, the inclusion of the similarity score function enhances the output quality for both models, further improving their overall performance.

\begin{table}[ht]
\centering
\begin{tabular}{cc}
\hline
Embedding         & Moving Motor Vehicle \\
\hline
Label             & $16.97$                \\
Motor             & $13.14$                \\
A moving motor      & $12.12$                \\
Sound of motor    & $12.87$                \\
Driving/motor/car & $12.07$                \\
Audio embedding  & $\textbf{8.88}$     \\
\hline
\end{tabular}
\caption{The results of motor sounds between different filter strategies, the embedding indicates the text value(Label is just a single number) for both training and calculating the similarity score. Embedding value with more than one means that the results need to pass the filter score of all the embedding targets. The results are evaluated on LDM large model.}
  \label{table:motor}
\end{table}

Although there have been significant improvements across most sound classes, we have observed that the generation quality of motor sounds does not show a significant decrease in FAD scores. This could be attributed to the fact that many motor sounds contain noise-like elements, making it challenging for CLAP to accurately identify and extract embeddings that align well with the corresponding texts. However, as demonstrated in Table~\ref{table:motor}, the utilization of different filter-based embeddings specifically for the motor class has resulted in a significant improvement in sound quality. By selecting a set of highly matched embeddings from the training dataset, our system achieves a FAD score of $8.88$ for motor sounds. Consequently, this method has been implemented as a final enhancement in the submitted system, ensuring more consistent and high-quality outputs for the motor sound class.

\section{CONCLUSION}
This technical report describes the system we submitted to the DCASE 2023 challenge task 7. Our system leverages the latest diffusion-based model and applied several technologies to improve the resulting quality. To achieve the best performance, our submit system consists of two models with the same structure but different sizes. The smaller LDM, build with a small-scale CLAP, is designed to generate sound for dog bark, footstep, gunshot, keyboard and sneeze cough. A larger LDM, accompanied by a bigger CLAP, focuses on synthesizing "moving motor" and "rain" sounds. The experimental result indicates that our system can significantly improve the baseline network by a large margin. 
\section{ACKNOWLEDGMENT}

This research was partly supported by a research scholarship from the China Scholarship Council~(CSC) No.$202208060240$, the British Broadcasting Corporation Research and Development~(BBC R\&D), Engineering and Physical Sciences Research Council~(EPSRC) Grant EP/T019751/1 ``AI for Sound'', and a PhD scholarship from the Centre for Vision, Speech and Signal Processing~(CVSSP), University of Surrey. For the purpose of open access, the authors have applied a Creative Commons Attribution~(CC BY) license to any Author Accepted Manuscript version arising.

\bibliographystyle{IEEEtran}

%
%
%
%
%
%
%
%
%

\end{sloppy}
\end{document}